\documentclass{iopart}
\usepackage{iopams}  
\usepackage[pdftex]{graphicx}
\usepackage{subfigure}
\usepackage{sidecap}

\def\qout      {q_{\rm out}}
\def\qside     {q_{\rm side}}
\def\qlong     {q_{\rm long}}

\def\Qout      {$\qout$}
\def\Qside     {$\qside$}
\def\Qlong     {$\qlong$}

\def\rout      {R_{\rm out}}
\def\rside     {R_{\rm side}}
\def\rlong     {R_{\rm long}}

\def\Rout      {$\rout$}
\def\Rside     {$\rside$}
\def\Rlong     {$\rlong$}

\def\ee        {$e^+e^-$}
\def\pbpb      {Pb--Pb}
\def\auau      {Au--Au}

\def\kt        {$k_{\rm T}$}
\def\meankt    {$\langle$\kt$\rangle$}

\def\sqrts     {$\sqrt{s_{\rm NN}}$}
\def\ene       {$\sqrt{s_{\rm NN}}=2.76$~TeV}
\def\dndeta    {${\rm d}N_{\rm ch}/{\rm d}\eta$}

\def\meandndeta{$\langle$\dndeta$\rangle$}

\def\gevc      {~GeV/$c$}
\def\dens      {charged-particle pseudorapidity density}

\begin{document}

\title[Two-pion Bose-Einstein correlations in \pbpb\ collisions with ALICE]{Two-pion 
Bose-Einstein correlations \\ in \pbpb\ collisions at 2.76 TeV with ALICE}

\author{J.~Mercado for the ALICE Collaboration}

%\address{Ruprecht-Karls-Universit\"at Heidelberg, 
\address{Universit\"at Heidelberg, 
Philosophenweg 12, 
69120 Heidelberg, Germany}

\ead{mercado@physi.uni-heidelberg.de}

\begin{abstract}
We present the first measurement of pion source radii in \pbpb~collisions 
at the LHC. The radii were obtained by analyzing the Bose-Einstein 
enhancement in two-pion correlation functions. Like at lower energies, 
the radii drop with increasing transverse momentum, indicating presence 
of collective expansion. In absolute terms, all three radii (\Rout, 
\Rside, \Rlong) are larger than at RHIC, roughly consistent with a 
linear scaling with the cube root of the particle multiplicity. 
\end{abstract}

%\submitto{\JPG}
%\maketitle

\section{Introduction}

The main object of study of ALICE (A Large Ion Collider Experiment) is
matter at extremely high energy density created in central collisions 
of heavy ions at the Large Hadron Collider (LHC)~\cite{Aamodt:2008zz,
Carminati:2004fp,Alessandro:2006yt}. 
Experimentally, the expansion rate and the spatial extent at decoupling of the
highly compressed strongly-interacting system created in these collisions
are accessible via intensity interferometry, a technique which exploits the 
Bose--Einstein enhancement of identical bosons emitted close by in phase-space. 
This approach, known as Hanbury Brown--Twiss analysis 
(HBT)~\cite{HBT1,Hanbury:1954wr}, has been successfully applied in 
\ee~\cite{Kittel:2001zw}, hadron--hadron and lepton--hadron~\cite{Alexander:2003ug}, 
and heavy-ion~\cite{Lisa:2005dd} collisions. 

\section{Data analysis}

The data were collected in 2010 during the first lead beam running period 
of the LHC. 
For the present analysis we have used $1.6\times10^4$ events that
correspond to the most central 5\% of the hadronic cross section. 
The \dens\ measured in this centrality class is 
\meandndeta~$=1601 \pm 60$~(syst.)~\cite{Collaboration:2010cz}.
The correlation analysis was performed using charged-particle tracks detected in 
the Inner Tracking System (ITS) and the Time Projection Chamber (TPC) of
ALICE. 

The two-particle correlation function is defined as the ratio 
$C\left({\bf q}\right)=A\left({\bf q}\right)/B\left({\bf q}\right)$, 
where  $A\left({\bf q}\right)$ is the measured distribution of the 
difference ${\bf q}={\bf p}_2-{\bf p}_1$ between the three-momenta of the two 
particles ${\bf p}_1$ and ${\bf p}_2$, 
and $B\left({\bf q}\right)$ is the corresponding distribution formed by using 
pairs of particles where each particle comes from a different event 
(event mixing).

\begin{figure}[!h]
\centering
\mbox{\subfigure{\includegraphics[width=0.38\textwidth]{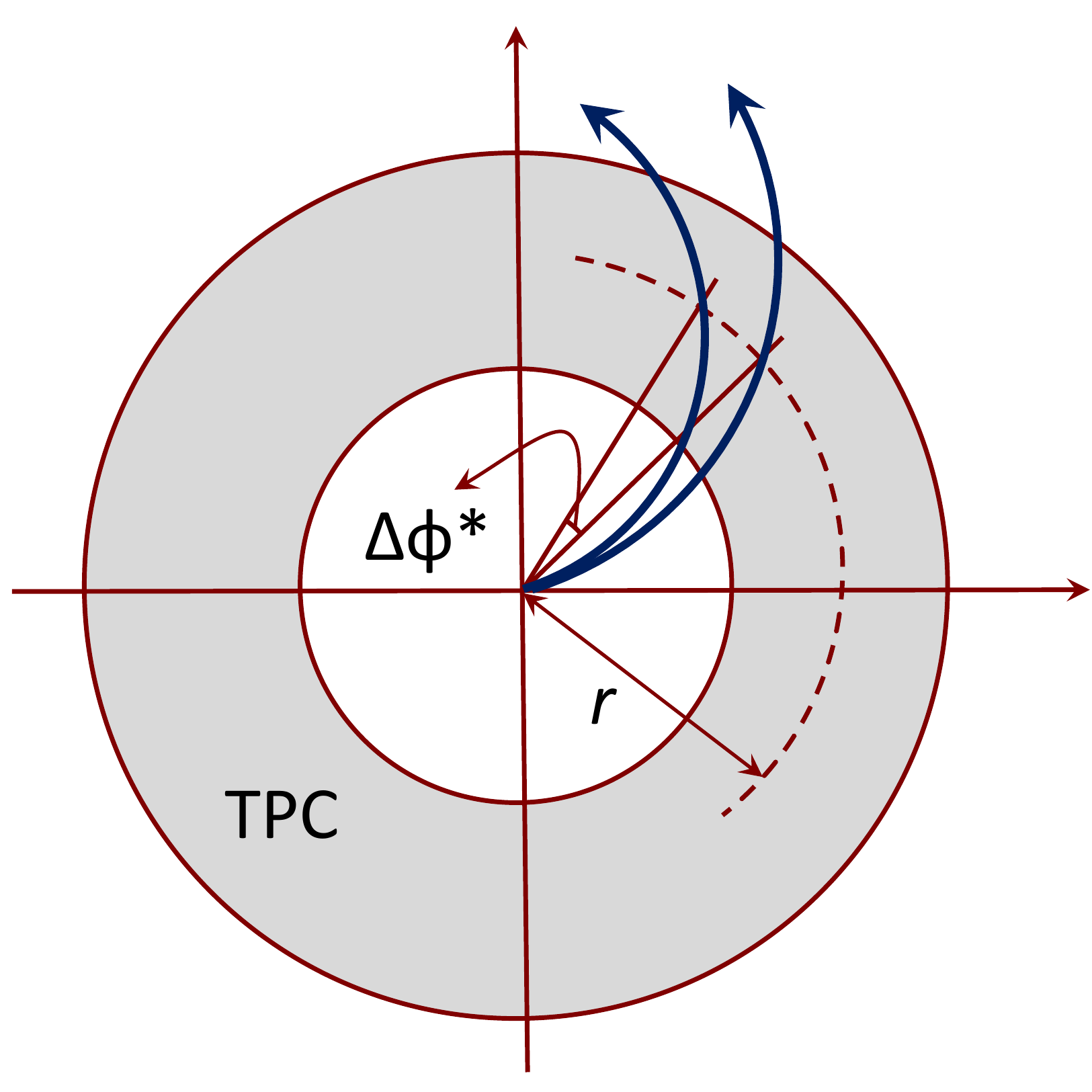}}\quad
\subfigure{\includegraphics[width=0.58\textwidth]{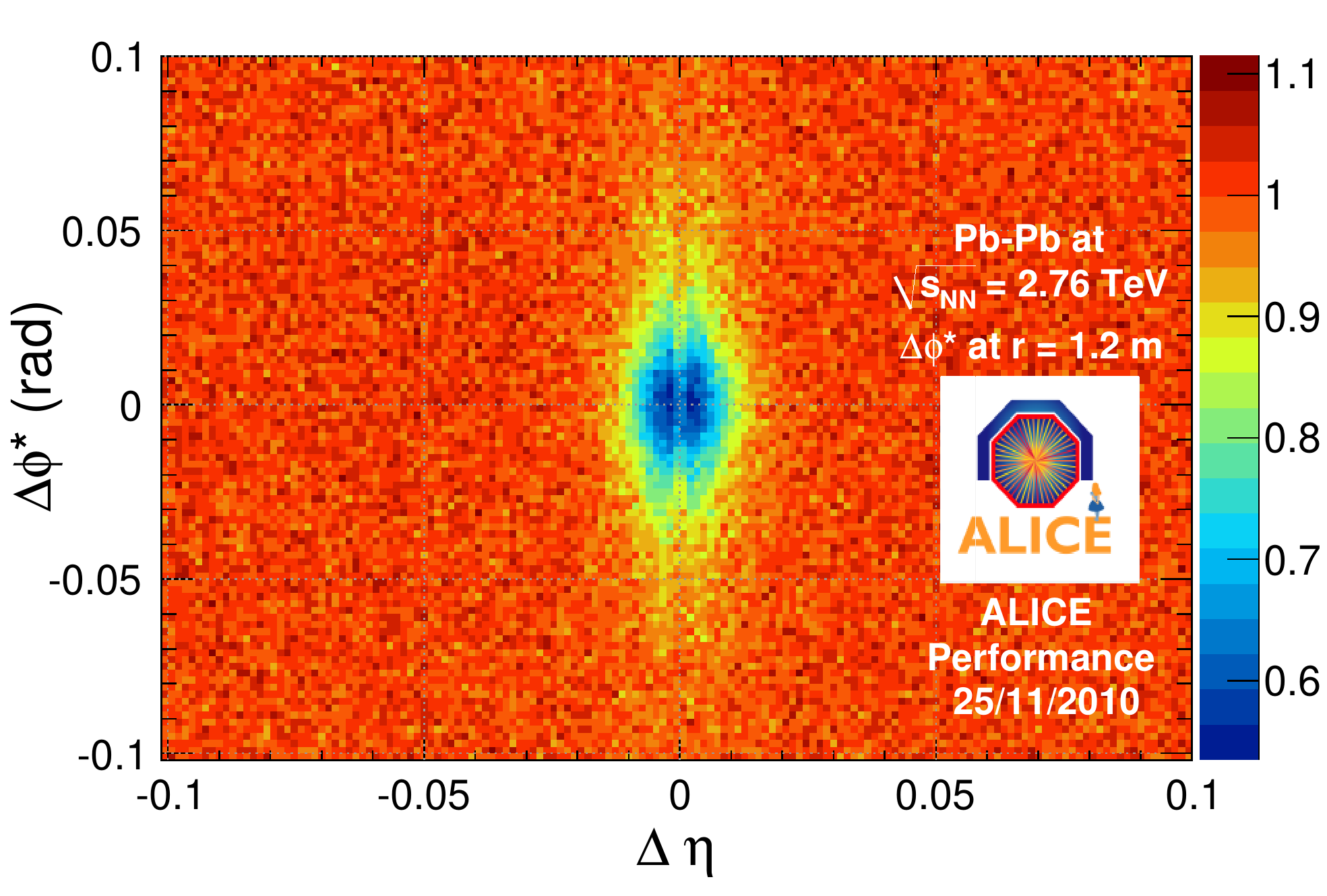}}}
\caption{Separation between tracks within the TPC (left). Pair 
inefficiency in the separation variable $\Delta\phi^{\star}$ calculated 
from momenta (right).} \label{ttr}
\end{figure}

The efficiency with which two tracks are reconstructed within the TPC is a 
function of the separation between them. 
Track splitting (incorrect reconstruction of a signal produced by one 
particle as two tracks) and track merging (reconstructing one track instead 
of two) generally lead to structures in the 
two-particle correlation functions if not properly treated. 
In this analysis, the track splitting effect is negligible
and the track merging leads to a 20-30\% loss of track pairs with a distance 
of closest approach in the TPC of 1~cm or less. 
This can be demonstrated by studying, in Monte Carlo simulations, 
distributions of reconstructed pairs in the $(\Delta\eta,\Delta\phi^{\star})$ 
plane. The two variables represent the longitudinal opening angle and the 
transverse angular separation at a given cylindrical radius, respectively,
and are defined as
\begin{eqnarray}
\Delta\eta &=& \eta_2 - \eta_1, \nonumber \\
\Delta\phi^{\star} &=& \Delta\varphi+
\arcsin(0.075\,r/p_{\rm{T,2}})
-\arcsin(0.075\,r/p_{\rm{T,1}}).
%\arcsin\left(\frac{0.075\cdot r}{p_{\rm{T,2}}}\right) 
%-\arcsin\left(\frac{0.075\cdot r}{p_{\rm{T,1}}}\right).
\end{eqnarray}
Here, $\Delta\varphi=\varphi_{2}-\varphi_{1}$ is the initial transverse opening
angle and $p_{\rm{T,1}}$ and $p_{\rm{T,2}}$ are the two transverse momenta in 
GeV$/c$\,; 
the cylindrical radius $r$ is in m. The pair inefficiency in the 
separation variable $\Delta\phi^{\star}$ is found to be sharpest at $r=1.2$~m
(Fig.~\ref{ttr}).
We avoid the two-track efficiency problem by including in the
correlation analysis only those pairs for which $|\Delta\phi^{\star}|<0.02$~rad or $|\Delta\eta|<0.01$.

The three-dimensional correlation functions $C$(\Qout, \Qside, \Qlong) 
were studied in bins of transverse momentum, \kt~$=|{\bf p}_{\rm{T,1}}+
{\bf p}_{\rm{T,2}}|/2$, from 0.2 to 1.0\gevc\ and fitted by an 
expression~\cite{Sinyukov:1998fc} accounting for the Bose--Einstein 
enhancement and for the Coulomb interaction between the two particles:
\begin{eqnarray}
\label{eq:1dbowler-sinyukov}
C({\rm \bf q})&=& \mathcal{N} \left\{ (1-\lambda)+\lambda K(q_{\rm inv}) [1 + G({\rm \bf q}) ] \right\} ,\\
\nonumber G({\rm \bf q}) &=& \exp[-(\rout^2 \qout^2 + \rside^2 \qside^2 + \rlong^2 \qlong^2 )],
\end{eqnarray}
with $\lambda$ describing the correlation strength, and \Rout, \Rside, and \Rlong\ 
being the Gaussian HBT radii. 
The factor $K(q_{\rm inv})$ is the squared Coulomb wave function 
averaged over a spherical source of size equal 
to the mean of \Rout, \Rside, and \Rlong.

\begin{figure}[!t]
\centering
\includegraphics[width=0.95\textwidth]{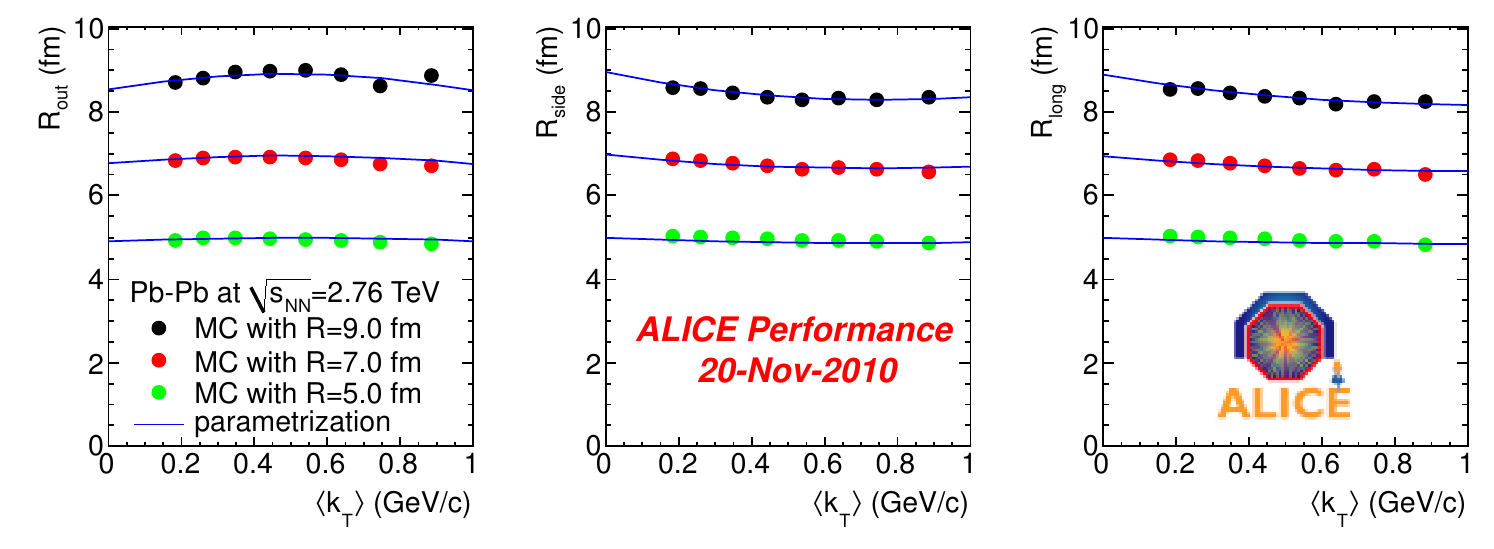}
\caption{Radii reconstructed in Monte Carlo with fixed radii
of 5, 7, and 9~fm, parametrized by a function approaching
unity in the limit of small radii.}  \label{fig:momres}
\end{figure}

The finite momentum resolution in the TPC smears out the correlation peak
causing the extracted radii to be smaller than they actually are. This effect
arises from the finite position resolution and from multiple scattering.
The effect was studied by applying weights to pairs of tracks in simulated 
events so as to produce the correlation function expected for a given 
set of the HBT radii (Fig.~\ref{fig:momres}). The reconstructed radii were found 
to differ from the input ones by up to 4\%, depending on the radius and \kt. 
The corresponding correction was applied to the experimental HBT radii. 

\section{Results}

\begin{figure}[!b]
\centering
\mbox{\subfigure{\includegraphics[width=0.55\textwidth]{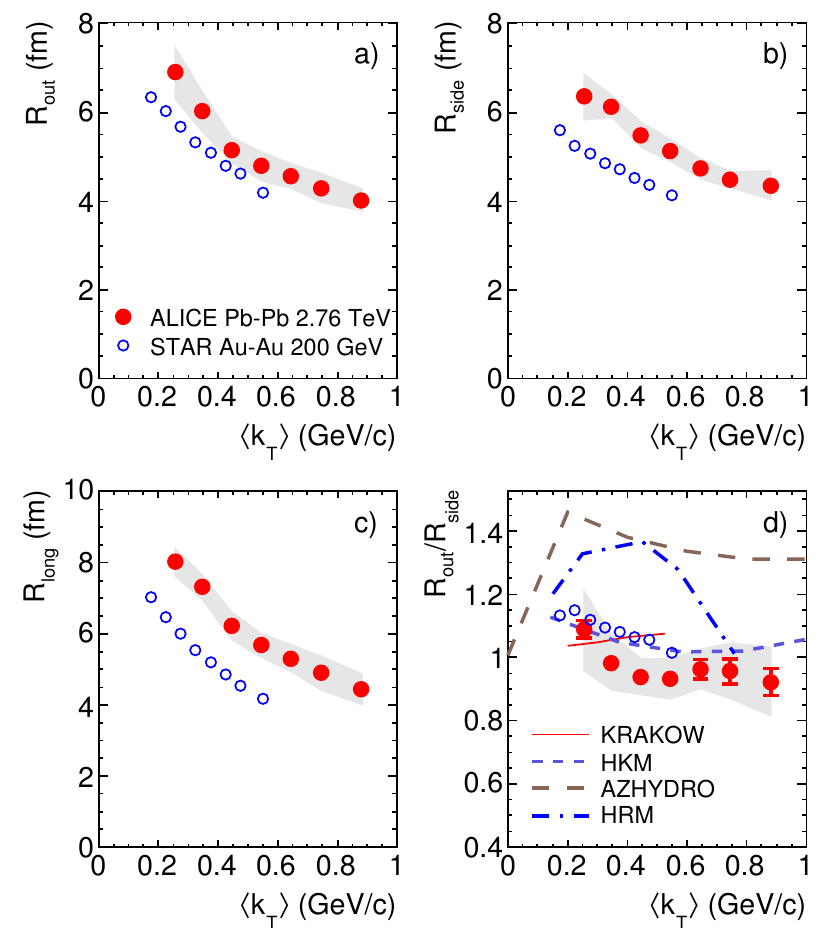}}\quad
\subfigure{\includegraphics[width=0.37\textwidth]{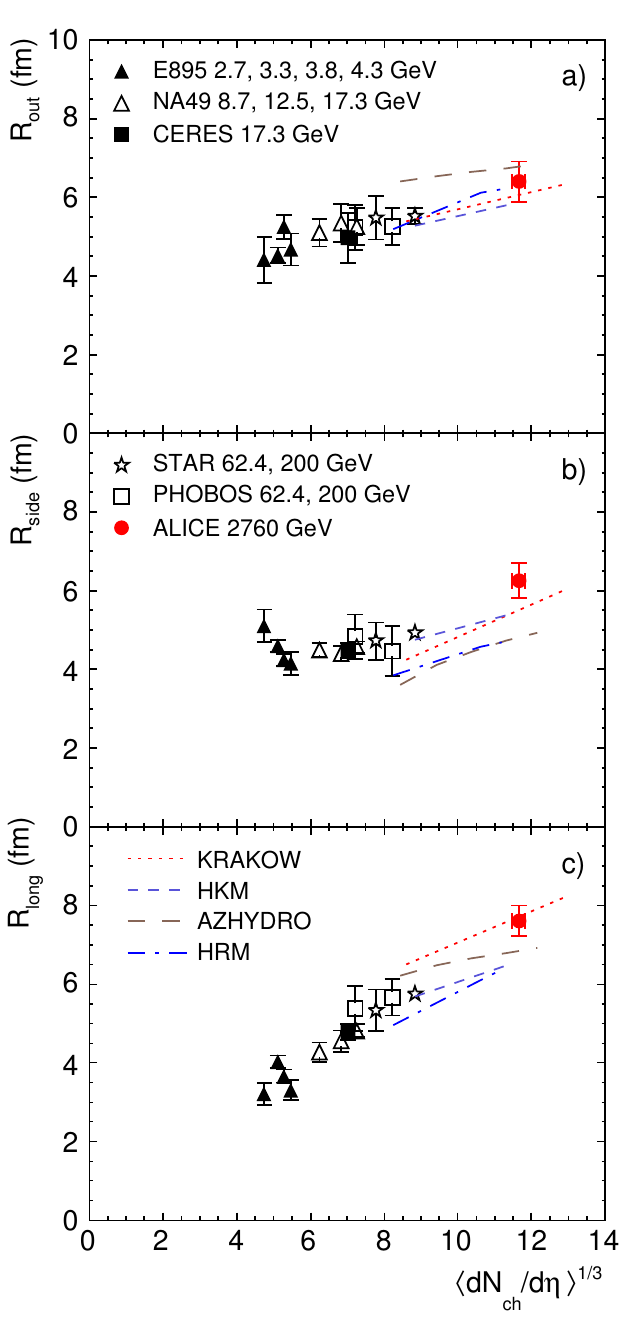}}}
\caption{Transverse momentum (left) and beam energy (right) dependence 
of the radii compared to those at lower energies. The lines show model 
predictions.} \label{fig:dep}
\end{figure}

The HBT radii extracted from the fit to the two-pion correlation functions 
and corrected as described in the previous section are shown as a function of 
\meankt\ in Fig.~\ref{fig:dep} (left)~\cite{Aamodt:2011pb}. 
The HBT radii for the 5\% most central \pbpb\ collisions at \ene\ 
are found to be significantly (10-35\%) larger than those measured by STAR in 
central \auau\ collisions at \sqrts~$=200$~GeV.
As also observed in heavy-ion collision experiments at lower 
energies, the HBT radii show a decreasing trend with increasing \kt\
which is a characteristic feature of expanding particle sources. 

The three radii at \meankt~$=0.3$~\gevc\ are compared with experimental results 
at lower energies in Fig.~\ref{fig:dep} (right). The trend of radii growing 
approximately linearly with the cube root of the \dens, established at lower 
energies, continues at the LHC.

Available model predictions are compared to the experimental data (Fig.~\ref{fig:dep}).
While the increase of the radii between RHIC and the LHC is roughly reproduced by 
all four calculations, only those tuned to reproduce RHIC data are able to describe 
the experimental \Rout/\Rside\ ratio as a function of \kt~(see Ref.~\cite{Aamodt:2011pb}). 

\begin{figure}[!h]
\centering
\mbox{\subfigure{\includegraphics[width=0.5\textwidth]{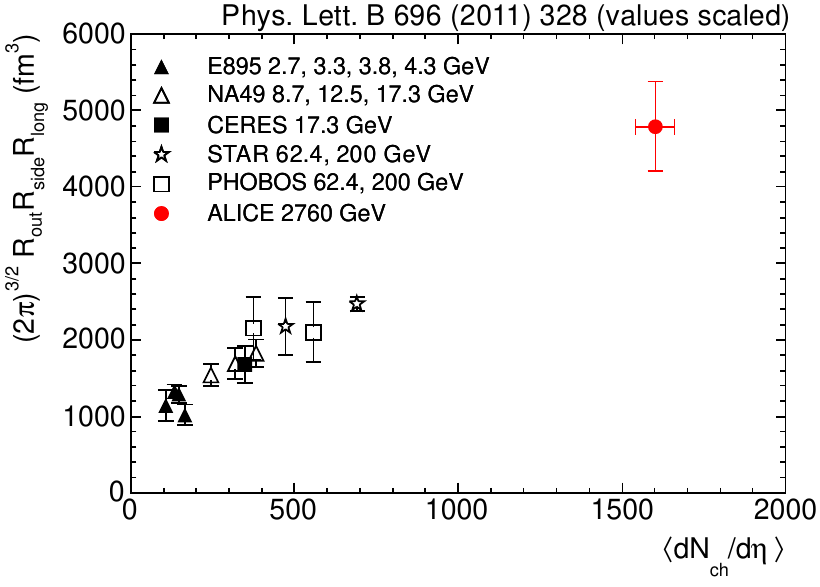}}\quad
\subfigure{\includegraphics[width=0.5\textwidth]{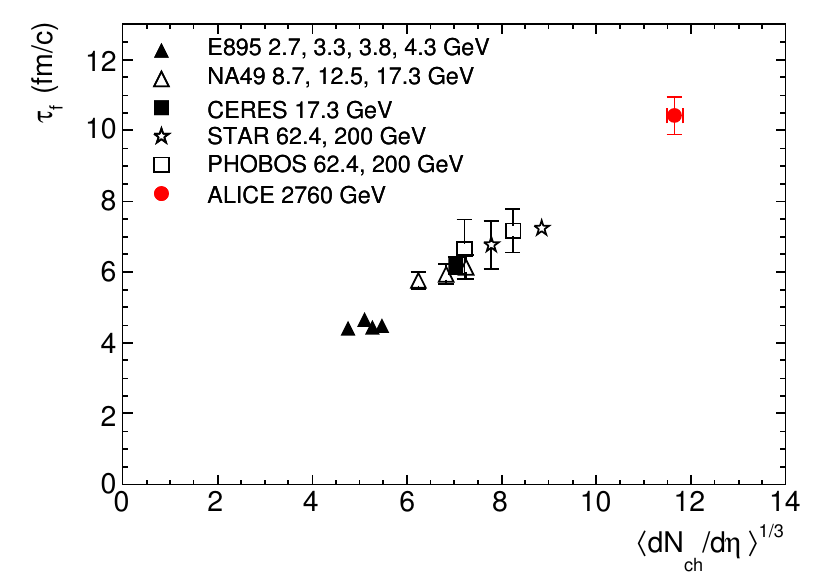}}}
\caption{Product of the three pion HBT radii (left) and decoupling time (right)
compared to the results obtained at lower energies.} \label{fig:size}
\end{figure}

The product of the three radii is related to the homogeneity volume and is 
found to be two times larger at LHC than at RHIC (Fig.~\ref{fig:size}, left).
Finally, the decoupling time extracted from \Rlong(\kt) for midrapidity pions
is 40\% larger than at RHIC (Fig.~\ref{fig:size}, right).

Our results, taken together with those obtained from the study of
multiplicity and the azimuthal anisotropy, indicate that the fireball
formed in nuclear collisions at the LHC is hotter, lives longer, and
expands to a larger size as compared to lower energies.

\end{document}